\documentclass[12pt]{iopart}
\usepackage[latin1]{inputenc} 
\usepackage[T1]{fontenc} 
\usepackage{textcomp,lmodern} 
\usepackage{amssymb, amsmath-modif} 
\DeclareMathOperator{\Spur}{Tr} %
\DeclareMathOperator{\diag}{diag}
\newcommand{\ens}[0]{\ensuremath} 
\newcommand{\anfEngl}[1]{``#1''} 
\newcommand{\ket}[1]{\ens{|#1\rangle}} 
\newcommand{\bra}[1]{\ens{\langle#1|}} 
\newcommand{\SkPr}[2]{\ens{\left\langle#1|#2\right\rangle}} 
\newcommand{\x}[0]{\ens{\otimes}} 
\newcommand{\Mge}[2]{\ens{\left\lbrace #1|\,#2 \right\rbrace}} 
\newcommand{\Mg}[1]{\ens{\left\lbrace #1 \right\rbrace}} 
\newcommand{\MgN}[1]{\ens{\Mg{0,\dots,#1}}} 
\newcommand{\MgZ}[1]{\ens{\Mg{2,\dots,#1}}} 
\newcommand{\betrag}[1]{\ens{|#1|}} 
\newcommand{\Eins}[0]{\ens{{\leavevmode{\rm 1\ifmmode\mkern -4.4mu\else\kern -.3em\fi I}}}} 
\newcommand{\pr}[1]{\ens{\ket{#1}\bra{#1}}} 
\newcommand{\Bew}[0]{\emph{Proof: }} 
\newcommand{\BE}[0]{\hfill $\Box$} 
\newcommand{\df}[1]{\mathrm{d}#1}

\renewcommand{\phi}[0]{\ens{\varphi}} 
\newcommand{\my}[0]{\ens{\mu}} 
\newcommand{\ny}[0]{\ens{\nu}} 
\newcommand{\ksi}[0]{\ens{\xi}}
\newcommand{\cH}[0]{\ens{\mathcal{H}}}
\newcommand{\fU}[0]{\ens{\mathfrak{U}}}
\newcommand{\N}[0]{\ens{\mathbb{N}}} 
\newcommand{\Z}[0]{\ens{\mathbb{Z}}}
\newcommand{\C}[0]{\ens{\mathbb{C}}}
\newcommand{\iE}[0]{\ens{\mathrm{i}}} 
\newcommand{\Alm}[0]{\ens{(A_{lm})_{l,m = 0}^{d-1}}}
\newcommand{\pers}[1]{#1} 
\newcommand{\zeit}[1]{\textit{#1}} 
\newcommand{\Band}[1]{\textbf{#1}} 
\newcommand{\Seiten}[2]{\mbox{#1--#2}} 
\newcommand{\ITIT}[0]{IEEE Trans. Inf. Theory}
\newcommand{\PRA}[0]{Phys. Rev. A}

\newtheorem{Definition}{Definition} 
\newtheorem{Lemma}{Lemma} 
\newtheorem{Theorem}{Theorem}
\newtheorem{Corollary}{Corollary} 

\begin{document}
\title{Symmetric extendibility for a class of qudit states}
\author{Kedar S Ranade}\ead{Kedar.Ranade@physik.tu-darmstadt.de}
\address{Institut f\"ur Angewandte Physik, Technische Universit\"at Darmstadt,\\
  Hochschulstraße 4a, D-64289 Darmstadt, Deutschland (Germany)}
\date{June 29, 2009}

\begin{abstract}
  The concept of symmetric extendibility has recently drawn attention in the context of tolerable error rates
  in quantum cryptography, where it can be used to decide whether quantum states shared between two parties
  can be purified by means of entanglement purification with one-way classical communication only. Unfortunately,
  at present there exists no simple general criterion to decide whether a state possesses a symmetric extension
  or not. In this article we derive criteria for symmetric extendibility within subclasses of all two-qudit
  states. Using these criteria, we can completely solve the problem for a two-parameter family of
  two-qudit states, which includes the isotropic states as a subclass.
\end{abstract}

\maketitle

\section{Introduction}
The concept of symmetric extendibility has recently been introduced into the field of quantum cryptography
as means to decide whether quantum states shared by two parties, Alice and Bob, may be purified by entanglement
purification protocols using one-way classical communication only. Whereas there exist criteria for the case
of two-qubit states which can be applied in quantum cryptography \cite{Myhr_ua,MyhrLuetkenhaus}, very little
is known about higher-dimensional states. The purpose of this work is to derive criteria for a subclass
of all two-qudit states, which may be applied in quantum cryptography using higher-dimensional quantum
systems (qudits) as carriers of information.
\par The outline of this article is the following: in this section we shall introduce the basic concepts and
notation; this includes the Hurwitz-Sylvester criterion for positivity, on which a large part of our discussion
relies. In section \ref{InvarStates} we introduce the class of $\fU_2$-invariant two-qudit states, which are
of interest in quantum cryptography \cite{RanadeAlber-2,Chau05}; for these states we derive a criterion
(Theorem \ref{SymErw}) in order to decide whether they are symmetrically extendible or not. We restrict our
focus to the class of Bell-diagonal $\fU_2$-invariant states, which are of even greater interest in quantum
cryptography \cite{RanadeAlber-2,Chau05,GottesmanLo} in section \ref{SecBell-diag} and simplify our criterion
to find Theorem \ref{SymErwBell}. In a subclass of these states we use this theorem to completely solve the
question of symmetric extendibility in a two-parameter family of two-qudit states, which form a superset of
the isotropic states. Finally, we conclude the paper with section \ref{SecConclusions}.

\subsection{Definition and basic facts}
We consider three $d$-dimensional Hilbert spaces $\cH_A = \cH_B = \cH_E = \C^d$, $d \in \N \setminus \Mg{1}$
(this naming arises from Alice, Bob and Eve in quantum cryptography),
each of which has a basis labelled by the elements of the ring of residue classes $\Z/d\Z$. This ring we shall
identify with the numbers in $\Z_d := \MgN{d-1}$, where all the operations (in particular, addition \anfEngl{$\oplus$}
and subtraction \anfEngl{$\ominus$}) are taken modulo $d$. In the following
we take a basis to be $\Mg{\ket{0},\,\ket{1},\,\dots,\ket{d-1}} \subseteq \C^d$ and all sums run over $\Z_d$.
We start with the definition of symmetric extendibility; in a more general context, it
may be called $(1,2)$-symmetric extendibility \cite{Terhal_ua}, but this is not within the scope of this work.
\begin{Definition}[Symmetric extendibility]\label{DefSymErw}\hfill\\
  A state $\rho_{AB}$ on $\cH_A \x \cH_B$ is called \emph{symmetrically extendible}, if
  there exists a state $\rho_{ABE}$ on $\cH_A \x \cH_B \x \cH_E$ with $\cH_E = \cH_B$, such that
  $\rho_{ABE} = \rho_{AEB}$ and $\Spur_E \rho_{ABE} = \rho_{AB}$ hold.
\end{Definition}
Obviously all separable states have a symmetric extension, whilst no pure entangled state does.
The general solution to the problem, whether a state is symmetrically extendible or not is unsolved, however,
a criterion for Bell-diagonal two-qubit states is known~\cite{Myhr_ua} and, more generally, criteria
for general two-qubit states have been investigated \cite{MyhrLuetkenhaus}.
\par To describe the problem more explicitly, consider two general density matrices on the Hilbert spaces
$\cH_A \x \cH_B$ and $\cH_A \x \cH_B \x \cH_E$, respectively:
\begin{eqnarray}
  \rho_{AB} &= \sum\nolimits_{ijpq} a_{ij,pq} \ket{ij}\bra{pq},\label{rho-AB}\\
  \rho_{ABE} &= \sum\nolimits_{ijkpqr} a_{ijk,pqr} \ket{ijk}\bra{pqr}.
\end{eqnarray}
In order for $\rho_{ABE}$ to be a symmetric extension of $\rho_{AB}$ three conditions must hold:
\begin{itemize}
  \item Symmetry (between $B$ and $E$): $a_{ijk,pqr} = a_{ikj,prq}$ for all $i,\,j,\,k,\,p,\,q,\,r \in \Z_d$;
  \item Trace condition (or extension property): $\sum_{k \in \Z_d} a_{ijk,pqk} = a_{ij,pq}$
    for all $i,\,j,\,p,\,q \in \Z_d$;
  \item Positivity (including hermiticity): $\rho_{ABE} \geq 0$.
\end{itemize}
The third property guarantees that $\rho_{ABE}$ is a quantum state, and the interplay between all three
conditions causes the main problem in determining whether a symmetric extension exists or not.

\subsection{The Hurwitz-Sylvester criterion}
For our purposes the most useful condition for checking, whether a matrix is positive (more precisely,
positive \emph{semidefinite}), is the Hurwitz-Sylvester criterion, which we will briefly explain in the following:
Let $A \in \C^{d \times d}$ be an arbitrary matrix represented
with respect to some fixed basis set, e.\,g. $B = \Mg{\ket{0},\,\ket{1},\,\dots,\ket{d-1}}$. Choosing any non-empty
subset $S \subseteq B$ with cardinality $r = \betrag{S}$, we can construct the associated $r \times r$ matrix
by skipping all rows and columns of $A$, whose basis vectors do not appear in $S$; the determinants of such subsets
are called \emph{principal minors} of order $r$, and there are altogether $2^d-1$ principal minors of $A$. We now
state the criterion; cf. e.\,g. \cite[p. 282]{Gantmacher}.
\begin{Lemma}[Hurwitz-Sylvester criterion for positivity]\label{Hurwitz}\hfill\\
  A matrix $A \in \C^{d \times d}$ is positive, if and only if all its principal minors are non-negative.
\end{Lemma}
This criterion is not to be confused with the better known \pers{Hurwitz}-\pers{Sylvester} criterion for positive
\emph{definite} matrices, which states that a matrix is positive definite, if and only if all \emph{leading}
principal minors, that is the determinants of the $d$ upper left submatrices, are (strictly) positive.
Note in particular that Lemma \ref{Hurwitz} implies that block-diagonal matrices are positive, if and
only if all blocks are positive.

\section{Symmetric extendibility of $\fU_2$-invariant states}\label{InvarStates}
In this section we introduce the class of states we are interested in, the $\fU_2$-invariant states. These
states were shown to be of interest in quantum cryptography \cite{RanadeAlber-2}, which is the main impetus
for our investigation. We will derive a criterion (Theorem \ref{SymErw}) in order to decide whether there
exists at least one possible symmetric extension.

\subsection{Invariant states and commutants}
It is yet not feasible to derive a criterion to decide whether an arbitrary two-qudit state possesses a
symmetric extension or not. Thus, in order to progress we have to choose an appropriate class of these
states, which should both be of physical interest and enable us to find a criterion for symmetric
extendibility. A convenient way of describing states is by their commutant. Consider for example the full
unitary group $\fU(\C^d \x \C^d)$ on the Hilbert space $\C^d \x \C^d$ of two qudits; we may ask which states
are \emph{invariant} with respect to that group. In this particular case Schur's lemma tells us that the
only invariant state is $d^{-2}\Eins_{d^2}$, since $\fU(\C^d \x \C^d)$ is irreducible. More interesting
examples are the states invariant with respect to $U \x U$ for all $U \in \fU(\C^d)$ (Werner states)
or with respect to $U \x U^*$ for all $U \in \fU(\C^d)$ (isotropic states).
\par In the following we shall focus on a superset of the set of the isotropic states. To this aim, let
us define three groups:
\begin{equation}\begin{array}{ll}
  \fU_1 &:= \Mge{U \in \fU(\C^d)}{U\,\text{diagonal in the standard basis}},\\
  \fU_2 &:= \Mge{U \x U^*}{U \in \fU_1},\\
  \fU_3 &:= \Mge{U \x U^* \x U^*}{U \in \fU_1}.
\end{array}\end{equation}
We may call $\fU_1$ the \emph{diagonal unitary group}; it is a maximally commutative subgroup of $\fU(\C^d)$,
and any matrix $U \in \fU_1$ may be written in the form $U = \diag(w_0,w_1,\dots,w_{d-1})$ for some system
$w = (w_0,\,w_1,\,\dots,w_{d-1}) \in \C^d$ of complex numbers which lie on the unit circle of $\C$.

\subsection{The class of $\fU_2$-invariant states}
The class of states we want to consider is the class of $\fU_2$-invariant states, which we describe now.
Given an arbitrary $U_w = \sum_{x = 0}^{d-1} w_x \pr{x} \in \fU_1$ and a two-qudit state in the form of (\ref{rho-AB}),
we calculate
\begin{equation}\begin{array}{lll}
  (U_w \x U_w^*)\rho_{AB} &= \sum_{xyijpq} w_x w_y^* a_{ij,pq} \ket{xy}\SkPr{xy}{ij}\bra{pq}
                           = \sum_{ijpq} w_i w_j^* a_{ij,pq} \ket{ij}\bra{pq},\\
  \rho_{AB}(U_w \x U_w^*) &= \sum_{xyijpq} w_x w_y^* a_{ij,pq} \ket{ij}\SkPr{pq}{xy}\bra{xy}
                           = \sum_{ijpq} w_p w_q^* a_{ij,pq} \ket{ij}\bra{pq},
\end{array}\end{equation}
and in order to be $\fU_2$-invariant, the two expressions have to be equal for all possible choices of $U_w$.
We thus have to ensure $w_i w_j^* a_{ijpq} = w_p w_q^* a_{ijpq}$ for all $i,\,j,\,p,\,q \in \Z_d$. If
$a_{ij,pq}$ is non-zero, this amounts to $w_i w_q = w_p w_j$, and since $U_w$ is arbitrary, this can be guaranteed
only if either $(i,q) = (j,p)$ or $(i,q) = (p,j)$ holds. Thus, all coefficients except those of the form
$a_{ii,pp}$ or $a_{ij,ij}$ must vanish, and the matrix is diagonal up to a block of size $d$ for the basis
vectors $\Mg{\ket{00},\,\ket{11},\,\dots,\ket{d-1,d-1}}$.

\subsection{The $\fU_3$-invariant states}
If it exists at all, a $\fU_2$-invariant state will have a $\fU_3$-invariant symmetric extension. This is
because for any symmetric extension $\rho_{ABE}$ of $\rho_{AB}$ and any $U \in \fU_3$, the state
$U\rho_{ABE}U^\dagger$ symmetrically extends $\rho_{AB}$. Averaging over the (unique) normalised Haar measure
on $\fU_3$ will yield the invariant extension $\rho_{ABE}^\prime = \int_{U \in \fU_3} U\rho_{ABE}U^\dagger \,\df{U}$.
Algebraically spoken, if there exists an extension, it can be chosen to lie in the commutant of $\fU_3$ in the
algebra of operators on $(\C^d)^{\x 3}$.
\par Since $\fU_3$ is commutative, it is easy to calculate its commutant, i.\,e. the $\fU_3$-invariant
states. This can be done in a similar fashion as we did for $\fU_2$ in the previous subsection, and we find
that $a_{ijk,pqr}$ may be non-zero, only if $(i,q,r)$ and $(p,j,k)$ are related by a permutation.
This leads to a block-matrix structure in the standard basis of $(\C^d)^{\x 3}$, which we can label by the
basis vectors; the blocks are
\begin{enumerate}
  \item blocks $B_k$ of size $2d-1$ for basis vectors $\ket{pkp}$ and $\ket{ppk}$ for $p \neq k$
    and $\ket{kkk}$,
  \item blocks $C_{ijk}$ of size $2$ for vectors $\ket{ijk}$ and $\ket{ikj}$, $i,\,j,\,k$ being all different,
  \item blocks $D_{ij}$ of size $1$ for the vector $\ket{ijj}$ with $i \neq j$.
\end{enumerate}
To recall our previous statements, given any extension of our state, we find an extension by setting all
elements to zero, which do not lie in any of these blocks. By using the block structure it gets much easier
to check positivity (see the note below Lemma \ref{Hurwitz}).

\subsection{The trace conditions}\label{Trace}
Any two-qudit state can be written as $\rho_{AB} = \sum_{ij,pq} a_{ij,pq} \ket{ij}\bra{pq}$; an extension
will then have the form $\rho_{ABE} = \sum_{ijk,pqr} a_{ijk,pqr} \ket{ijk}\bra{pqr}$, and we have to
determine the coefficients $a_{ijk,pqr}$. In the case $k = r$ they have to obey certain trace conditions,
and we want to check where these coefficients $a_{ijk,pqk}$ lie. We consider the two cases of nonzero coefficients
of $\rho_{AB}$:
\begin{enumerate}
  \item $a_{ii,pp}$: the relevant coefficients $a_{iik,ppk}$ lie in the blocks $B_k$;
  \item $a_{ij,ij}$: the relevant coefficients $a_{ijk,ijk}$ are the diagonal elements of all blocks.
\end{enumerate}
The remaining coefficients $a_{ij,pq}$ are zero due to the $\fU_2$-invariance, and we set $a_{ijk,pqk} := 0$,
since they lie outside of our block structure. We note that the off-diagonal elements $a_{ijk,ikj}$ and $a_{ikj,ijk}$
of $C_{ijk}$ can be set to zero, since they do not appear in the trace and according to Lemma \ref{Hurwitz}
any other choice may only harm positivity of $\rho_{ABE}$.

\subsection{Symmetry and the reduction of $B_k$ to $B_k^\prime$}
Apart from the trace condition we still have to fulfil the symmetry $a_{ijk,pqr} = a_{ikj,prq}$. In the case
of the blocks $D_{ij}$ nothing has to be done, and for $C_{ijk}$ we note that it is a multiple of the
$2 \times 2$ unit matrix. Let us therefore focus on the blocks $B_k$.
\par Each block $B_k$ is constructed for the basis vectors $\ket{ppk}$ and $\ket{pkp}$ for $k \neq p$ and the
exceptional element $\ket{kkk}$. By symmetry $a_{iik,ppk} = a_{iki,pkp}$ and $a_{iik,pkp} = a_{iki,ppk}$ hold; whilst
the first-mentioned elements appear in the trace condition, the latter do not. We now choose
$a_{iik,pkp} := a_{iik,ppk}$ and show that this is not a restriction. Let $B_k^\prime$ be the $d \times d$
submatrix of $B_k$ constructed for the basis vectors $\ket{ppk}$ (where $k = p$ is possible).
\begin{Lemma}[Equivalence of positivity of $B_k$ and $B_k^\prime$]\hfill\\
  Either $B_k$ and $B_k^\prime$ are both positive semidefinite or none of them is.
\end{Lemma}
\Bew If $B_k$ is positive definite, then so is its submatrix $B_k^\prime$. Assuming that $B_k^\prime$ is positive
semidefinite, we choose an arbitrary principal minor of $B_k$. If is is constructed by using a pair
$\ket{ppk}$ and $\ket{pkp}$, it is zero due to our choice of the elements $a_{iik,pkp}$; if not, we can replace
all $\ket{pkp}$ by $\ket{ppk}$ to yield a submatrix of $B_k^\prime$. Positivity is thus ensured by
Lemma \ref{Hurwitz}. \BE\vspace{0.2cm}\\
Since the elements $a_{iik,pkp}$ do not appear in $B_k^\prime$, any other choice may only harm positivity.
Furthermore, by this reduction, we got rid of the symmetry constraint, which is now implicitly hidden in the
matrices.

\subsection{Building up the matrices $B_k^\prime$}
We now want to explicitly construct positive matrices $B_k^\prime$. For shortness, let us denote
$\lambda_{ijk} := a_{ijk,ijk}$ and $\lambda_{ij} := a_{ij,ij}$ for the diagonal elements; the symmetry and
the second trace condition then read $\lambda_{ijk} = \lambda_{ikj}$ and $\sum_k \lambda_{ijk} = \lambda_{ij}$.
For fixed $i \in \Z_d$, we can write a scheme, which is symmetric and consists of non-negative entries:
\begin{center}\begin{tabular}{c|cccccc|c}
    \begin{tabular}{cc} $k$ : column index\\ $j$ : row index\end{tabular}
             & $0$             & $1$             & $\hdots$      & $i$             & $\hdots$ & $d-1$
                                                                 & row sum\\
    \hline
    $0$      & $\lambda_{i00}$ & $\lambda_{i01}$ &  $\hdots$     & $\lambda_{i0i}$ & $\hdots$ & $\lambda_{i,0,d-1}$
                                                                                   &  $\lambda_{i0}$\\
    $1$      & $\lambda_{i10}$ & $\lambda_{i11}$ &  $\hdots$     & $\lambda_{i1i}$ & $\hdots$ & $\lambda_{i,1,d-1}$
                                                                                   & $\lambda_{i1}$\\
    $\vdots$ & $\vdots$        & $\vdots$        & $\ddots$      & $\vdots$        &          &  $\vdots$    & $\vdots$\\
    $i$      & $\lambda_{ii0}$ & $\lambda_{ii1}$ & $\hdots$      & $\lambda_{iii}$ & $\hdots$ & $\lambda_{i,i,d-1}$
                                                                                   & $\lambda_{ii}$\\
    $\vdots$ & $\vdots$        &  $\vdots$       &       & $\vdots$ & $\ddots$  & $\vdots$    & $\vdots$\\
    $d-1$    & $\lambda_{i,d-1,0}$ & $\lambda_{i,d-1,1}$ & $\hdots$ & $\lambda_{i,d-1,i}$ & $\hdots$ & $\lambda_{i,d-1,d-1}$
                                                                                  & $\lambda_{i,d-1}$\\
    \hline column sum
             & $\lambda_{i0}$ & $\lambda_{i1}$ & $\hdots$ & $\lambda_{ii}$ & $\hdots$ & $\lambda_{i,d-1}$ &
\end{tabular}.\end{center}
The elements on the \anfEngl{cross} defined by $i = j$ or $i = k$ lie in the blocks $B_k$, the remaining diagonal
entries in blocks $D_{ij}$ and all other in blocks $C_{ijk}$. The second trace condition fixes the sum of each
row and each column.
\par Given such a scheme, positivity has to be ensured within the blocks $B_k^\prime$ only. If there exists
a scheme which fulfils all criteria and produces positive $B_k^\prime$, there exists a scheme, where the
$C_{ijk}$ vanish: if some $\lambda_{ijk} =: x \geq 0$, by symmetry $\lambda_{ikj} = x$
holds. Substituting $\lambda_{ijj}^\prime := \lambda_{ijj} + x$, $\lambda_{ikk}^\prime := \lambda_{ikk} + x$
and $\lambda_{ijk}^\prime := \lambda_{ikj}^\prime := 0$, the trace conditions are still fulfilled, $C_{ijk} = 0$
and the diagonal elements of the $B_k^\prime$ remain unaffected.
\par We can thus arbitrarily choose the diagonal entries of the matrices $B_k^\prime$ between zero and its
maximum value, since the $D_{ij}$, i.\,e. the entries $\lambda_{ijj} := \lambda_{ij} - \lambda_{iij}$
will absorb the remaining value to fulfil the trace condition. The only thing we have to take care of is
$\lambda_{iik} \leq \lambda_{ik}$ for all $i,\,k \in \Z_d$, since the first trace condition ensures
$\sum\nolimits_{p \in \Z_d} \lambda_{ppk} = \lambda_{pp}$ for all $k \in \Z_d$.

\subsection{Reformulation of the trace condition and the main theorem}
The matrix $B_k^\prime$ is constructed with respect to the basis vectors $\ket{ppk}$ for $p \in \Z_d$, where
we now consider this particular ordering. Summing up all matrices $B_k^\prime$ yields
\begin{equation}
  \sum\nolimits_{k = 0}^{d-1} B_k^\prime = \left(\sum\nolimits_k a_{iik,ppk}\right)_{i,p=0}^{d-1}
    = \left(a_{ii,pp}\right)_{i,p=0}^{d-1} =: \tilde{B}
\end{equation}
according to the first trace condition, and as a submatrix of $\rho_{AB}$, it is always positive.
Skipping the primes in $B_k^\prime$, we have altogether shown the following theorem.
\begin{Theorem}[Symmetric extendibility of $\fU_2$-invariant states]\label{SymErw}\hfill\\
  A $\fU_2$-invariant state $\rho_{AB} = \sum_{ijpq} a_{ij,pq} \ket{ij}\bra{pq}$ is symmetrically extendible,
  if and only if the matrix $\tilde{B} = (a_{ii,pp})_{i,p=0}^{d-1} \in \C^{d \times d}$ can be decomposed into the
  sum of $d$ positive matrices $B_k = \left(a_{iik,ppk}\right)_{i,p=0}^{d-1} \in \C^{d \times d}$ for
  $k \in \Z_d$, such that their diagonal elements obey the inequalities
  $a_{iik,iik} \leq a_{ik,ik}$ for all $i,\,k \in \Z_d$.
\end{Theorem}
In general, this condition is still difficult to check, however, it is sufficiently appropriate for calculating
bounds for quantum-cryptographic protocols \cite{TeilB}, and we will use it as a starting point for the next section.
\par Since the sum of positive matrices is positive, we can always enlarge the diagonal elements of a positive
matrix without changing its positivity. Ignoring for the moment the trace conditions,
we could set the diagonal elements of all $B_k$ to their maximum values. Considering only the
non-negativity of all principal minors constructed of $2 \times 2$ submatrices, we find the following
corollary.
\begin{Corollary}[Necessary condition for symmetric extendibility]\label{Corollary}\hfill\\
  A $\fU_2$-invariant symmetrically extendible state fulfils
  $\betrag{a_{ii,pp}} \leq \sum_{k = 0}^{d-1} \sqrt{a_{ik,ik} a_{pk,pk}}$ for all $i,\,p \in \Z_d$.
\end{Corollary}

\section{Bell-diagonal states}\label{SecBell-diag}
An important subset of all two-qudit states is the class of (generalised) Bell-diagonal states. We define
the Bell basis of the Hilbert space $\cH = \C^d \x \C^d$ by
\begin{equation}
  \ket{\Psi_{lm}} := d^{-1/2} \sum\nolimits_{k = 0}^{d-1} z^{lk} \ket{k} \ket{k \ominus m},\quad l,\,m \in \Z_d,
\end{equation}
where $z := \exp\big(\frac{2\pi\iE}{d}\bigr)$ is the principal value of the $d$-th root of unity.
The Bell-diagonal states are the convex combinations of the associated density matrices and can be written
in the form
\begin{equation}\label{Bell-diag}
  \rho_{AB} = \sum\nolimits_{l,m=0}^{d-1} A_{lm} \pr{\Psi_{lm}},
\end{equation}
where $A_{lm} \geq 0$ and $\sum_{lm} A_{lm} = 1$. The coefficient system $\Alm$ thus defines a probability
distribution, and we write $A_{*m} := \sum_{l = 0}^{d-1} A_{lm}$ for one of its marginals.
To construct the elements $a_{ij,pq}$, we rewrite (\ref{Bell-diag}) as
\begin{equation}
  \rho = d^{-1} \sum\nolimits_{lmkk^\prime} A_{lm} z^{l(k-k^\prime)} \ket{k,k\ominus m}\bra{k^\prime,k^\prime\ominus m}
\end{equation}
and thus find $a_{ij,pq} = d^{-1} \delta_{i \ominus j,p \ominus q} \sum_l A_{l,i \ominus j} z^{l(i-p)}$; since
$\delta_{i \ominus j,p \ominus q} = \delta_{i \ominus p,j \ominus q}$, this gives rise to a block structure
of the density matrix, where for every $m \in \Z_d$ the basis elements of the blocks are given
by $\Mge{\ket{ip}}{i \ominus p = m}$.
Comparing this with the block structure of general $\fU_2$-invariant states, we find the following Lemma.
\begin{Lemma}[Characterisation of $\fU_2$-invariant Bell-diagonal states]\hfill\\
  A Bell-diagonal state with coefficient system $\Alm$ is $\fU_2$-invariant, if and only if for all $m \neq 0$
  and $l \in \Z_d$ there holds $A_{lm} = d^{-1}A_{*m}$.
\end{Lemma}
The two trace conditions of subsection \ref{Trace} now read
\begin{equation}\label{Spurbed}
  \sum_{k} a_{ijk,pqk} \stackrel{!}{=} a_{ij,pq} = \begin{cases}
        d^{-1}\tilde{A}_{ip} := d^{-1}\sum_{l} A_{l0} z^{l(i-p)}, & \text{if $i = j$ and $p = q$}\\
        d^{-1}A_{*,i \ominus j} = \lambda_{ij}, & \text{if $i = p$ and $j = q$}.
      \end{cases}
\end{equation}
Note that there is no ambiguity in the case $i = j = p = q$, and the remaining cases are all zero and irrelevant.
As in subsection \ref{Trace}, the relevant components for the first trace condition lie in the blocks $B_k$,
whilst the relevant components for the second trace condition are precisely the diagonal elements
of all blocks.

\subsection{Symmetric extensions of $\fU_2$-invariant Bell-diagonal states}
The Bell-diagonal states have particular properties, which we can use in our discussion. Namely, the matrix
$\tilde{B} = d^{-1} (\tilde{A}_{ip})_{i,p=0}^{d-1}$ of Theorem \ref{SymErw} is circulant and the conditions
on the diagonal elements of the $B_k$ also have the circulant structure $\lambda_{iik} \leq d^{-1} A_{*,i \ominus k}$.
This will yield some simplifications.
\par The symmetric group $S_d$ can be seen to consist of the permutations on $\Z_d$. Using a permutation $\pi \in S_d$,
one can shift rows and columns of a matrix $A = (a_{ij})_{i,j = 0}^{d-1} \in \C^{d \times d}$ to get
$A^{(\pi)} = (a_{\pi(i),\pi(j)})_{i,j = 0}^{d-1}$. (Technically spoken, this is a representation of $S_d$ on $\C^d$.)
For the cyclic permutation defined by $\pi_l(i) := i \ominus l$, we shall write $A^{(l)} := A^{(\pi_l)}$.
With this definition we can simplify Theorem \ref{SymErw} in the case of Bell-diagonal states.
\begin{Theorem}[Symmetric extendibility of $\fU_2$-invariant Bell-diagonal states]\label{SymErwBell}\hfill\\
  For a $\fU_2$-invariant Bell-diagonal symmetrically extendible state, the set of matrices in Theorem \ref{SymErw}
  can be chosen to consist of matrices $B_0,\,B_1,\,\dots,B_{d-1}$, such that $B_l = B_0^{(l)}$ holds for all
  $l \in \Z_d$.
\end{Theorem}
\Bew First note that in the Bell-diagonal case, the matrix $\tilde{B}$ of Theorem \ref{SymErw} is circulant in
the Bell-diagonal case, i.\,e. $\tilde{B} = \tilde{B}^{(l)}$
for all $l \in \Z_d$. This implies
\begin{equation}
  \tilde{B} = \tilde{B}^{(l)} = B_0^{(l)} + B_1^{(l)} + B_2^{(l)} + \dots + B_{d-1}^{(l)},
\end{equation}
and we can define $B_k^\prime := d^{-1} \sum_{l = 0}^{d-1} B_{k \ominus l}^{(l)}$ for all $k \in \Z_d$. Since
the matrix $B_k^{(l)}$ fulfils the same diagonal constraints as $B_{k \oplus l}$, the matrix $B_k^\prime$
fulfils the same conditions as $B_k$, and $\sum_{k = 0}^{d-1} B_k^\prime = \tilde{B}$. \BE\vspace{0.2cm}\\
This Theorem tells us, that we effectively have to look for \emph{one} matrix $B_0$ only instead of $d$ matrices.
Corollary \ref{Corollary} now states that symmetrically extendible $\fU_2$-invariant Bell-diagonal states fulfil
$\betrag{\tilde{A}_{ip}} \leq \sum_{k = 0}^{d-1} \sqrt{A_{*k} A_{*,k \oplus i \ominus p}}$ for all $i,\,p \in \Z_d$.

\subsection{Generalised-isotropic states}\enlargethispage{\baselineskip}
We now want to concentrate on an even more restricted class of states, where we can solve the problem completely,
the \emph{generalised isotropic states} \cite{RanadeAlber-2}. These are Bell-diagonal states where
$A_{l0} = A_{l^\prime0}$, $A_{0m} = A_{0m^\prime}$ and $A_{lm} = A_{l^\prime m^\prime}$ hold for all $l,\,m \neq 0$.
Since we enforce $\fU_2$-invariance and normalisation, we are left with two parameters, $a$ and $b$ only, for which
there hold $a,\,b \geq 0$ and $x := a+(d-1)b \leq 1$; we have
\begin{equation}\label{v-iso}
  A_{lm} = \begin{cases}
             a, & \text{if $l = m = 0$,}\\
             b, & \text{if $l \neq m = 0$,}\\
             \frac{1-a-(d-1)b}{d(d-1)} & \text{else.}
           \end{cases}
\end{equation}
In particular, $A_{*m} = \delta_{m0} \cdot x + (1-\delta_{m0}) \cdot \frac{1-x}{d-1}$ and
$\sum_{l} A_{l0} z^{l(i-p)} =  \delta_{ip} \cdot x + (1-\delta_{ip})(a-b)$. For the moment, we exclude the
case $d = 2$ due to some notational complications, but will discuss it later on.
Considering the matrix $B_0^\prime$ of Theorem \ref{SymErwBell}, the constraints on the diagonal elements read
$a_{000,000} \leq d^{-1} \cdot x$ and $a_{ii0,ii0} \leq d^{-1} \cdot \frac{1-x}{d-1}$ for $i \neq 0$. We shall
now consider the matrix $B_0^{\prime\prime}$, where we average all rows and columns except the first one:
\begin{equation}
  B_0^{\prime\prime} := \frac{1}{(d-1)!} \sum\nolimits_{\pi \in \Mge{\phi \in S_d}{\phi(0) = 0}} B_0^{\prime(\pi)}.
\end{equation}
A positive sum of positive matrices being positive, the matrix $B_0^{\prime\prime}$ is positive and can replace
$B_0^\prime$ in Theorem \ref{SymErwBell}, because the sums of the off-diagonal components are the same as
in $B_0^\prime$, as is shown in the following. The use of this mixing over several permutations enforces some
symmetries; we write $B_0^\prime = (b_{ij})_{i,j = 0}^{d-1}$ for $b_{ij} := a_{ii0,jj0}$:
\begin{enumerate}
  \item The entry $b_{00}$ remains unaffected and invariant,
  \item the entries $b_{0j}$, $j \neq 0$, are mapped to $(d-1)^{-1}(b_{01} + b_{02} + \dots + b_{0,d-1})$,
  \item the entries $b_{i0}$, $i \neq 0$, are mapped to $(d-1)^{-1}(b_{10} + b_{20} + \dots + b_{d-1,0})$,
  \item the entries $b_{ij}$, $i = j \neq 0$, are mapped to $(d-1)^{-1}(b_{11} + b_{22} + \dots + b_{d-1,d-1})$,
  \item the entries $b_{ij}$, $i \neq j$, $i,\,j \neq 0$, are mapped to $(d-1)^{-1}(d-2)^{-1}
    \sum_{i \neq j,\,i,\,j \neq 0} b_{ij}$.
\end{enumerate}
We can thus focus on matrices of the form $B_0^{\prime\prime} = d^{-1}M_d(\alpha,\beta,\ksi,\eta)$, where
\begin{equation}
  M_d(\alpha,\beta,\ksi,\eta) := \begin{pmatrix}
        \alpha & \ksi^* & \ksi^* & \hdots & \ksi^* \\
        \ksi   & \beta  & \eta   & \hdots & \eta \\
        \ksi   & \eta   & \beta  & \ddots & \vdots \\
        \vdots & \vdots & \ddots & \ddots & \eta \\
        \ksi   & \eta   & \hdots & \eta   & \beta
      \end{pmatrix} \in \C^{d \times d};
\end{equation}
the determinant of this matrix is given by
\begin{equation}\label{Det}
  \det M_d(\alpha,\beta,\ksi,\eta)
    = (\beta-\eta)^{d-2} \Bigl\{ \alpha \bigl[\beta + (d-2)\eta\bigr] - (d-1) \betrag{\ksi}^2 \Bigr\}.
\end{equation}
In order for $B_0^{\prime\prime}$ to be hermitian, $\alpha$, $\beta$ and $\eta$ must be real; the parameter
$\ksi$ can be chosen to be real, since $\ksi + \ksi^* + (d-2)\eta \stackrel{!}{=} a-b$ is real, and replacing
$\ksi$ by its real part $\mathrm{Re}\,\ksi$ does not change the sum and does not harm positivity of the matrix,
which will be a consequence of the following Lemma.
\begin{Lemma}[Positive semidefinite matrices]\hfill\\
  The matrix $M_d(\alpha,\beta,\ksi,\eta)$  is positive semidefinite, if and only if the three quantities
  $\alpha$, $\beta$ and $\det M_d(\alpha,\beta,\ksi,\eta)$ are jointly non-negative, the inequality
  $\betrag{\ksi} \leq \sqrt{\alpha\beta}$ holds and $\eta \in [-\frac{\beta}{d-2};\beta]$.
\end{Lemma}
\Bew Using Lemma \ref{Hurwitz}, we have to check whether all principal minors of $M_d(\alpha,\beta,\ksi,\eta)$
are non-negative. The principal minors of order one are $\alpha$, $\beta$, the others can easily seen to be
\begin{equation}\begin{array}{ll}
  \det M_r(\alpha,\beta,\ksi,\eta) & \text{for $r \in \MgZ{d}$},\\
  \det M_s(\beta,\beta,\eta,\eta) & \text{for $s \in \MgZ{d-1}$}.
\end{array}\end{equation}
By invoking (\ref{Det}) we find $\det M_s(\beta,\beta,\eta,\eta) = (\beta-\eta)^{s-1}\bigl[\beta+(s-1)\eta\bigr]$,
which leads to $\eta \in [-\frac{\beta}{d-2};\beta]$. For $\det M_r(\alpha,\beta,\ksi,\eta)$ we thus
focus on the curly bracket of $(\ref{Det})$ to find $(\beta-\eta)^{-(r-1)} \det M_{r+1}(\alpha,\beta,\ksi,\eta)
= (\beta-\eta)^{-(r-2)} \det M_r(\alpha,\beta,\ksi,\eta) + (\alpha\eta - \betrag{\ksi}^2)$. Since
$(\alpha\eta - \betrag{\ksi}^2)$ is fixed, we only need to consider the cases $r \in \Mg{2,d}$, which are
given by $\betrag{\ksi} \leq \sqrt{\alpha\beta}$ and $\det M_d(\alpha,\beta,\ksi,\eta) \geq 0$,
respectively. \BE\vspace{0.2cm}\\
Let us for now denote by $\rho(a,b)$ the state described by (\ref{v-iso}), which is the general form of an
$\fU_2$-invariant Bell-diagonal generalised-isotropic state. To satisfy Theorem~\ref{SymErwBell},
$\alpha + (d-1)\beta = a + (d-1)b = x$ must hold. For $x$ being fixed, we may thus write $\alpha = (1-\sigma)x$
and $\beta = \frac{\sigma x}{d-1}$, where the diagonal constraints from Theorem \ref{SymErw} read
$\sigma \in [0;\min\Mg{1, \frac{1-x}{x}}]$. The following Lemma allows us to focus on the extremal values
$(a-b)_{\min} \leq 0 \leq (a-b)_{\max}$ for which the state is symmetrically extendible, given that $x$ is fixed.
\begin{Lemma}[Mixtures of states]\hfill\\
  Given two symmetrically extendible states $\rho(a_1,b_1)$ and $\rho(a_2,b_2)$, such that there holds
  $x = a_1 + (d-1)b_1 = a_2 + (d-1)b_2$, any other state $\rho(a,b)$ with $a + (d-1)b = x$
  and $a_1 - b_1 \leq a - b \leq a_2 - b_2$ is symmetrically extendible.
\end{Lemma}
\Bew We find $\rho(a,b) = p \cdot \rho(a_1,b_1) + (1-p) \cdot \rho(a_2,b_2)$ for
$p := \frac{(a-b)-(a_2-b_2)}{(a_1-b_1)-(a_2-b_2)}$ and note that the set of symmetrically extendible states
is convex. \BE\vspace{0.2cm}\\
We will now investigate the possible choices of $\ksi$ and $\eta$ to find the allowed values for
$2\ksi + (d-2)\eta = a - b$.

\subsubsection{Calculation of $(a-b)_{\max}$}\label{ab-max}\hfill\\
To find $(a-b)_{\max}$, it is sufficient to maximise $\ksi$ and $\eta$ individually. We can therefore set
$\eta_{\max} := \beta = \frac{\sigma x}{d-1}$, which leads to the maximum range for $\ksi$. The determinant
condition $\det M_d(\alpha,\beta,\ksi,\eta) \geq 0$ leads to
\begin{equation}
  \betrag{\ksi} \leq \sqrt{\frac{\alpha \bigl[\beta + (d-2)\eta_{\max}\bigr]}{d-1}}
                = \sqrt{\frac{(1-\sigma)x \bigl[\frac{\sigma x}{d-1} + (d-2)\frac{\sigma x}{d-1}\bigr]}{d-1}}
                = x \sqrt{\frac{\sigma(1-\sigma)}{d-1}},
\end{equation}
which is precisely the same as the other condition $\ksi \leq \sqrt{\alpha\beta}$. We have thus found
$\ksi_{\max} = x \sqrt{\frac{\sigma(1-\sigma)}{d-1}}$, which results in
$(a-b)_{\max} = 2\ksi_{\max} + (d-2)\eta_{\max} = x \cdot f(\sigma)$ for
\begin{equation}
  f(\sigma) := 2 \cdot \sqrt{\frac{\sigma(1-\sigma)}{d-1}} + (d-2) \cdot \frac{\sigma}{d-1},
\end{equation}
and we still have to maximise over $\sigma \in [0;\min\Mg{1, \frac{1-x}{x}}]$. The function $f$ monotonically
increases up to a maximum value of $f(\frac{d-1}{d}) = 1$. If the choice of $\sigma := \frac{d-1}{d}$ is allowed,
any state with positive $(a-b)$ is symmetrically extendible, since $a-b \leq x$ is always true; this holds,
if $\frac{d-1}{d} \leq \frac{1-x}{x}$ or $x \leq \frac{d}{2d-1}$. Else we choose the maximally possible value
$\sigma := \frac{1-x}{x}$ to find
\begin{equation}\label{abKrit}
  a - b \leq f\left(\frac{1-x}{x}\right) \cdot x = 2 \sqrt{\frac{(1-x)(2x-1)}{d-1}} + \frac{d-2}{d-1} \cdot (1-x)
\end{equation}
as a criterion for symmetric extendibility, given that $a - b \geq 0$.

\subsubsection{Calculation of $(a-b)_{\min}$}\label{ab-min}\hfill\\
The calculation of $(a-b)_{\min}$ is more involved than the previous one, because we cannot separately minimise
$\ksi$ and $\eta$. We write $\eta = \tau x$ and start with the conditions on $\ksi$:
\begin{equation}
  \betrag{\ksi} \leq \sqrt{\frac{\alpha \bigl[\beta + (d-2)\eta\bigr]}{d-1}}
    = x \sqrt{\frac{(1-\sigma) \bigl[\frac{\sigma}{d-1} + (d-2)\tau \bigr]}{d-1}}.
\end{equation}
Since $\eta \in [\frac{-\beta}{d-2};\beta]$, there must hold $\tau \in [\frac{-\sigma}{(d-2)(d-1)};\frac{\sigma}{d-1}]$.
We can continue to substitute $\my := (d-2)(d-1)\tau$ and $\ny := \my + \sigma$ to find
\begin{equation}
  \betrag{\ksi} \leq \frac{x}{d-1} \sqrt{(1-\sigma) (\sigma + \my)} = \frac{x}{d-1} \sqrt{(1-\sigma)\ny}
\end{equation}
for $\my \in [-\sigma;(d-2)\sigma]$ and $\ny \in [0;(d-1)\sigma]$. We can set
$\ksi_{\min} := -\frac{x}{d-1}\sqrt{(1-\sigma)\ny}$ and minimise the value of
\begin{equation}
  2\ksi + (d-2)\eta = \frac{x}{d-1} \cdot \left[-2 \sqrt{(1-\sigma)\ny} + (\ny-\sigma)\right].
\end{equation}
The first derivative of the bracket with respect to $\ny$ is $1 - \sqrt{\ny^{-1}(1-\sigma)}$, unless
$\sigma = 1$ or $\ny = 0$, and the minimum always lies in $[0;(d-1)\sigma]$. The minimum attained is $-1$, so
$(a-b)_{\min} = -\frac{x}{d-1}$, and since smaller values of $(a-b)$ are impossible by definition, \emph{all}
states with $a < b$ can be symmetrically extended.

\subsection{Discussion of results}
Altogether, the last two calculations of $(a-b)_{\max}$ and $(a-b)_{\min}$ have shown the following.
\begin{Theorem}[Symmetric extendibility of generalised-isotropic states]\label{SymErw-viso}\hfill\\
  For $d \geq 3$, an $\fU_2$-invariant Bell-diagonal generalised-isotropic state is symmetrically
  extendible, if and only if either $x \leq \frac{d}{2d-1}$ or inequality (\ref{abKrit}) or both hold.
\end{Theorem}
We shall finally discuss the qubit case $d = 2$. The calculations from \ref{ab-max} essentially go through,
but those of \ref{ab-min} fail due to denominators $d-2$. However, by a local unitary operation we can
interchange $a$ and $b$ and find that a state with $x \in [1/3;2/3]$ or
\begin{equation}
  \betrag{a - b} \leq 2 \sqrt{(1-x)(2x-1)}
\end{equation}
is symmetrically extendible; rewriting this yields $-9a^2-14ab-9b^2+12a+12b-4 \geq 0$, which coincides with
the results known before \cite{Myhr_ua}.
\par Another important case for two-qudit states are isotropic states (cf. subsection~\ref{InvarStates}). It can be
shown  that the isotropic states are those Bell-diagonal states where the equality $A_{lm} = \frac{1-A_{00}}{d^2-1}$ holds
for all $(l,m) \neq (0,0)$. In this case only a single parameter
is left (any one of $a$, $b$ or $x$). We find $(a-b)_{\mathrm{isotropic}} = \frac{dx-1}{d-1} > 0$ and by
solving (\ref{abKrit}) we find $x \leq \frac{d+3}{2(d+1)}$ or $a = x - (d-1)\frac{1-x}{d(d-1)} \leq \frac{d+1}{2d}$
to be necessary and sufficient for symmetric extendibility, if $d \geq 3$; for $d = 2$, the condition is
$a \in [1/4;3/4]$.

\section{Conclusions}\label{SecConclusions}
We have derived a criterion for symmetric extendibility of $\fU_2$-invariant two-qudit states
in terms of a matrix decomposition (Theorem \ref{SymErw}). We have simplified this in the case of Bell-diagonal
states (Theorem \ref{SymErwBell}), and for the two-parameter family of generalised-isotropic $\fU_2$-invariant states,
we have completely solved the problem (Theorem \ref{SymErw-viso}). The relevance of these three criteria
is shown by the fact, that particular instances can be used to derive bounds on tolerable error rates in
quantum cryptography \cite{TeilB,Ranade}.

\ack The author thanks \pers{Gernot Alber}, \pers{Matthias Christandl}, \pers{Norbert Lütkenhaus},
\pers{Geir Ove Myhr} and \pers{Joseph M. Renes} for helpful discussions. He was supported by a
\emph{Promotions\-stipendium} of the TU Darmstadt; financial support by CASED is acknowledged.


\section*{References}
\begin{thebibliography}{99}
  \bibitem{Myhr_ua}
    Myhr G O, Renes J M, Doherty A C and Lütkenhaus N 2009 \zeit{\PRA} \Band{79} 042329
  \bibitem{MyhrLuetkenhaus}
    Myhr G O and Lütkenhaus N 2008 \zeit{Preprint} arXiv:0812.3667v1
  \bibitem{RanadeAlber-2}
    Ranade K S and Alber G 2007 \zeit{J. Phys. A: Math. Theor.} \Band{40} \Seiten{139}{153}
  \bibitem{Chau05}
    Chau H F 2005 \zeit{\ITIT} \Band{51} \Seiten{1451}{1468}
  \bibitem{GottesmanLo}
    Gottesman D and Lo H-K 2003 \zeit{\ITIT} \Band{49} \Seiten{457}{475}
  \bibitem{Terhal_ua}
    Terhal B M, Doherty A C and Schwab D 2003 \zeit{\PRL} \Band{90} 157903
  \bibitem{Gantmacher}
    Gantmacher F R 1958 \textit{Matrizenrechnung} Teil I, VEB Deutscher Verlag der Wissenschaften (Berlin)
  \bibitem{TeilB}
    Ranade K S 2009 \zeit{Preprint} arXiv:0906.0163v1
  \bibitem{Ranade}
    Ranade K S 2009 Quantenkryptographie in endlichdimensionalen Systemen \textit{PhD thesis}\\
    \texttt{http://tuprints.ulb.tu-darmstadt.de/1318/1/Dissertation\_160\_2009-02-20\_modif2.pdf}
\end{thebibliography}
\end{document}